\documentclass[aps,prd,amsmath,preprint]{revtex4}

\usepackage{enumerate}
\usepackage{amsmath,amssymb}
\usepackage{bm}
\usepackage{amscd}            
\usepackage{graphicx}
\usepackage{hhline,multirow}  
\usepackage{dcolumn}          
\usepackage{url}              

\preprint{JLAB-THY-13-5}

\begin{document}
\title{Global parton distributions with nuclear and finite-$Q^2$ corrections}

\author{J.F.~Owens$^1$,
        A.~Accardi$^{2,3}$,
	W.~Melnitchouk$^3$}
\affiliation{
$^1$\mbox{Florida State University, Tallahassee, Florida 32306-4350}\\
$^2$\mbox{Hampton University, Hampton, Virginia 23668}	 \\
$^3$\mbox{Jefferson Lab, Newport News, Virginia 23606}
}

\date{\today}

\begin{abstract}
We present three new sets of next-to-leading order parton distribution
functions (PDFs) determined by global fits to a wide variety of data
for hard scattering processes. The analysis includes target mass and
higher twist corrections needed for the description of deep inelastic
scattering data at large $x$ and low $Q^2$, and nuclear corrections
for deuterium targets.
The PDF sets correspond to three different models for the nuclear
effects, and provide a more realistic uncertainty range for the $d$
quark PDF, in particular, compared with previous fits.  We describe
the PDF error sets for each choice of the nuclear corrections, and
provide a user interface for utilizing the distributions.
\end{abstract}

\maketitle

\section{Introduction}

Parton distribution functions (PDFs) are necessary ingredients for
any calculation of a hard scattering process involving hadrons in
the initial state.  Deeply-inelastic lepton-hadron scattering (DIS)
experiments provide data which yield information on PDFs over a wide
range of parton light-cone momentum fraction $x$ and four-momentum
transfer squared $Q^2$.  Traditional global fits focus on the
extraction of the leading twist PDFs, utilizing cuts on $Q^2$ and
the hadronic final state mass squared $W^2 = M^2 + Q^2 (1-x)/x$,
where $M$ is the nucleon mass.  These cuts, which typically are of
the order $Q^2 \gtrsim 4$~GeV$^2$ and $W^2 \gtrsim 14$~GeV$^2$, are
designed to eliminate regions where effects that do not scale with
$Q^2$ may be important.
At any finite value of $Q^2$, the available range in $x$ is therefore
kinematically limited by $x < x_{\rm max} = Q^2/(W^2 - M^2 + Q^2)$.
Available DIS data sets have few points above $x \approx 0.7$ when
these types of cuts are applied.  Other types of data such as lepton
pair production or high-$p_T$ jet production are also generally
limited to regions which are sensitive to PDF momentum fraction
values $x \lesssim 0.7$, or have limited statistics.  There are few
constraints, therefore, on the large-$x$ behavior of PDFs, and the
results of global fits using these cuts should be considered as
extrapolations when evaluated above this value of $x$.

Aside from the intrinsic interest in the behavior of PDFs at large
values of $x$ \cite{Feyn72, Close73, FJ75, MT96, HR10}, there are
also important practical aspects to consider.  The production of
a state of mass $m$ at rapidity $y$ in proton-proton collisions at
a center-of-mass energy $\sqrt{s}$ requires knowledge of PDFs at
$x \approx (m/\sqrt{s})\, e^{\pm y}$ at a scale of order $m$.
High mass states produced in the forward scattering region at
large positive values of $y$ therefore entail the product of one
PDF evaluated at a large value of $x$ with another at a small $x$.
Furthermore, PDFs at large $x$ and small factorization scale $Q$
evolve via the QCD evolution equations to lower $x$ and higher $Q$,
so that large-$x$ uncertainties in fixed-target experiments
can have significant consequences for collider measurements
\cite{Kuh00, Bra12}.

In an effort to obtain constraints on PDFs at large values of $x$,
a study was made by Accardi {\it et al.} \cite{CJ10} in which the
$Q^2$ and $W^2$ cuts applied to DIS data were systematically relaxed,
thereby allowing data at higher $x$ and lower $Q^2$ to be used.
This necessitated the inclusion of various corrections subleading
in $1/Q^2$, such as target mass corrections and higher twist
contributions, which become increasingly important as $Q^2$
is lowered and $x$ tends to 1.

A goal of global PDF fits is to determine the $x$ and $Q^2$ dependence
of the different parton species (gluons, individual quark and antiquark
flavors).  Charged-lepton DIS experiments on proton targets are sensitive
at large $x$ to the combination $4u +d$, but to separate the $u$ and $d$
PDFs an additional constraint is needed.  Data on a free neutron target
at large values of $x$ would constrain $4d+ u$; however, the closest
one can come to this is to utilize a deuterium target.
Since the proton and neutron in deuterium are not free, it is necessary 
to account for nuclear effects such as Fermi motion, binding, and
nucleon off-shell corrections, which can be significant at large values
of $x$.  Estimates of the uncertainties due to different models of these
nuclear corrections were studied in Ref.~\cite{CJ11}.

The purpose of the present paper is to synthesize the lessons of the
previous studies \cite{CJ10, CJ11}, and present three sets of global
next-to-leading order (NLO) PDF fits, each of which differs by the size
of the nuclear corrections employed for the deuteron target DIS data.
The results are referred to collectively as the CJ12 PDFs (with ``CJ''
denoting the CTEQ-Jefferson Lab collaboration \cite{CJweb}), and the
three PDF sets with mild, medium and strong nuclear corrections are
designated as ``CJ12min'', ``CJ12mid'' and ``CJ12max'', respectively.
These PDFs provide a convenient way to explore the effects of the
nuclear corrections on various observables.

In the following section we describe the methodology involved in the
fits, including the choices of data sets, the parametrizations used,
and the treatment of nuclear and finite-$Q^2$ corrections.  We also
discuss the treatment of the experimental errors and the resulting
error PDF sets.  In Sec.~\ref{sec:results} we present an overview of
the CJ12 PDF results, and compare these with results from other global
fits.  Finally, in Sec.~\ref{sec:conclusion} we make some concluding
remarks and outline plans for future work.  An appendix is provided
which contains the initial parameter values for each of the PDF sets.

\section{Formalism and procedures} 
\label{sec:formalism}

The CJ12 PDF sets described herein result from global fits to a wide
variety of data totaling 3958 data points.  The global fitting
procedures utilize NLO calculations for each observable.  For the DIS
observables, higher twist contributions and target mass corrections
are included, and for the deuterium DIS data sets nuclear corrections
are taken into account, as described in this section.

\subsection{Data}

The data sets used, listed in Table~\ref{tab:datasets}, encompass data
ranging from medium energy electron-hadron scattering at Jefferson Lab
to high-energy hadron-hadron scattering at the Fermilab Tevatron.
The kinematic cuts employed in this analysis,
$Q^2 > Q_0^2 = (1.3\ \rm GeV)^2$ and
$W^2 > 3$~GeV$^2$, allow the inclusion of: $F_2$ structure function
measurements on hydrogen and deuterium from
  BCDMS \cite{BCDMS},
  SLAC \cite{SLAC},
  NMC \cite{NMCp, NMCdop}, and
  Jefferson Lab \cite{Malace};
neutral current (NC) and charged current (CC) DIS cross sections
  from HERA \cite{HERA1};
Drell-Yan data from Fermilab \cite{E866};
$W$ (and decay lepton) asymmetries from CDF and D\O\
  \cite{CDF98, CDF05, D008, D0_e08, CDF09};
$Z$ rapidity distributions from CDF and D\O\ \cite{CDFZ, D0Z};
jet production cross sections from both CDF \cite{CDFjet1, CDFjet2},
  and D\O\ \cite{D0jet1, D0jet2}, and photon plus jet production from
  D\O\ \cite{D0gamjet}.

DIS data with a deuterium target are necessary for $u$ and $d$ quark
flavor separation at large $x$, which entails the use of nuclear
corrections, as will be discussed and evaluated below.
Unlike global fits from other groups, the present data set does not 
include the NuTeV dimuon data \cite{Mason07} from neutrino-iron
scattering, which provide essentially the only currently available
source of information on the $s$ and $\bar{s}$ quark distributions
(except possibly for $W$ and $Z$ production at the LHC \cite{ATLAS-s}).
The reason is that, beside corrections due to modifications of
medium-$x$ quark PDFs in the heavy nuclear target, which are generally
not too large \cite{nCTEQ09, nCTEQ11, Eskola, DSZS, Kumano},
the uncertainty due to in-medium propagation of the tagged final
state charm quark and $D$ meson may be significant \cite{Acc10}.
Similarly, we do not use E605 data on dimuon production from a
copper target \cite{E605}, and rely on E866 data \cite{E866} only
to constrain antiquark distributions.

\begin{table}[tb]
\caption{Data sets and number of data points used in the global fits,
	together with the $\chi^2$ values for each of the CJ12min,
	CJ12mid and CJ12max fits.\\}
\centering
{\scriptsize 
\begin{tabular}[c]{llcc|ccc}  \hline
           &&&&\multicolumn{3}{c}{$\chi^2$} \\
           & Experiment & Ref. & \# points\ \
		&\ CJ12min &  CJ12mid & CJ12max\\
           \hline
DIS $F_2$  & BCDMS $(p)$ &\cite{BCDMS}      & 351 & 434 & 436 & 437 \\
           & BCDMS $(d)$ &\cite{BCDMS}      & 254 & 294 & 297 & 302 \\
           & NMC $(p)$   &\cite{NMCp}       & 275 & 434 & 432 & 430 \\
           & NMC $(d/p)$ &\cite{NMCdop}     & 189 & 179 & 177 & 182 \\
           & SLAC $(p)$  &\cite{SLAC}       & 565 & 456 & 455 & 456 \\
           & SLAC $(d)$  &\cite{SLAC}       & 582 & 394 & 388 & 396 \\
           & JLab $(p)$  &\cite{Malace}     & 136 & 170 & 169 & 170 \\
           & JLab $(d)$  &\cite{Malace}     & 136 & 124 & 125 & 126 \\
DIS $\sigma$&HERA (NC $e^-$) &\cite{HERA1}  & 145 & 117 & 117 & 118 \\
           & HERA (NC $e^+$) &\cite{HERA1}  & 384 & 595 & 596 & 596 \\
           & HERA (CC $e^-$) &\cite{HERA1}  & 34  & 19  & 19  & 19  \\
           & HERA (CC $e^+$) &\cite{HERA1}  & 34  & 32  & 32  & 32  \\
Drell-Yan  & E866 $(p)$ &\cite{E866}        & 184 & 220 & 221 & 221 \\
           & E866 $(d)$ &\cite{E866}        & 191 & 297 & 307 & 306 \\
$W$ asymmetry& CDF 1998 ($\ell$) &\cite{CDF98} & 11 & 14 & 16 & 18 \\  
           & CDF 2005 ($\ell$) &\cite{CDF05}   & 11 & 11 & 11 & 10 \\  
           & D\O\ 2008 ($\ell$)&\cite{D008}    & 10 & 4  & 4  & 4  \\  
           & D\O\ 2008 ($e$)   &\cite{D0_e08}  & 12 & 40 & 36 & 34 \\
           & CDF 2009 ($W$)    &\cite{CDF09}   & 13 & 20 & 25 & 41 \\
$Z$ rapidity & CDF ($Z$)       &\cite{CDFZ}    & 28 & 29 & 27 & 27 \\
           & D\O\ ($Z$)        &\cite{D0Z}     & 28 & 16 & 16 & 16 \\
jet        & CDF run 1         &\cite{CDFjet1} & 33 & 52 & 52 & 52 \\
           & CDF run 2         &\cite{CDFjet2} & 72 & 14 & 14 & 14 \\
           & D\O\ run 1        &\cite{D0jet1}  & 90 & 21 & 20 & 19 \\
           & D\O\ run 2        &\cite{D0jet2}  & 90 & 19 & 19 & 20 \\
$\gamma$+jet& D\O\ 1           &\cite{D0gamjet}& 16 & 6  & 6  & 6  \\
            & D\O\ 2           &\cite{D0gamjet}& 16 & 13 & 13 & 12 \\
            & D\O\ 3           &\cite{D0gamjet}& 12 & 17 & 17 & 17 \\
            & D\O\ 4           &\cite{D0gamjet}& 12 & 17 & 16 & 17 \\
\hline
\multicolumn{3}{c}{TOTAL}                      &3958&4059&4055&4096 \\
\multicolumn{3}{c}{TOTAL + norm}               &    &4075&4074&4117 \\
\hline\\
\end{tabular}
}
\label{tab:datasets}
\end{table}

\begin{figure}
\includegraphics[width=0.49\linewidth,trim= 0 100 0 10,clip=true]
                {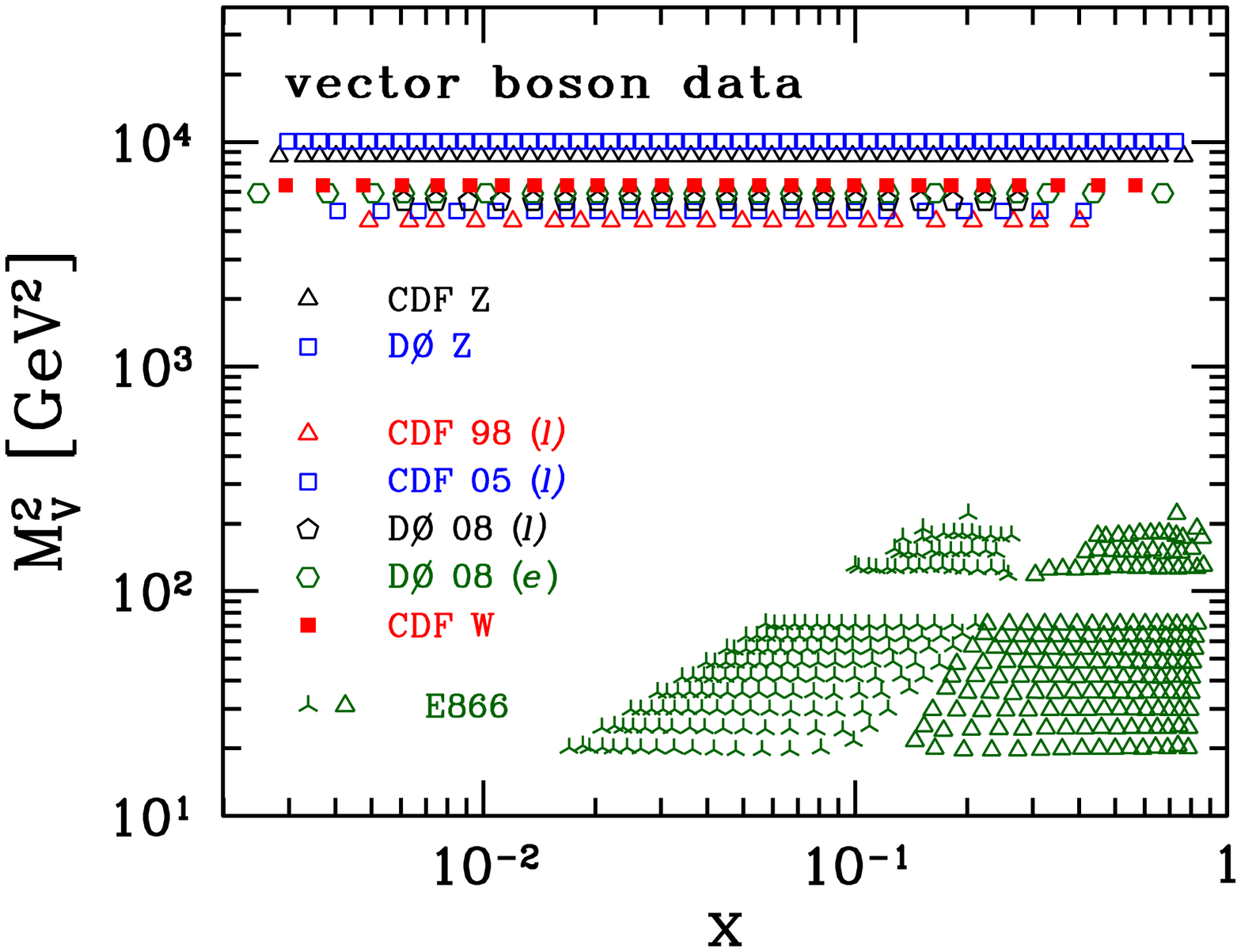}
\includegraphics[width=0.49\linewidth,trim= 0 100 0 10,clip=true]
                {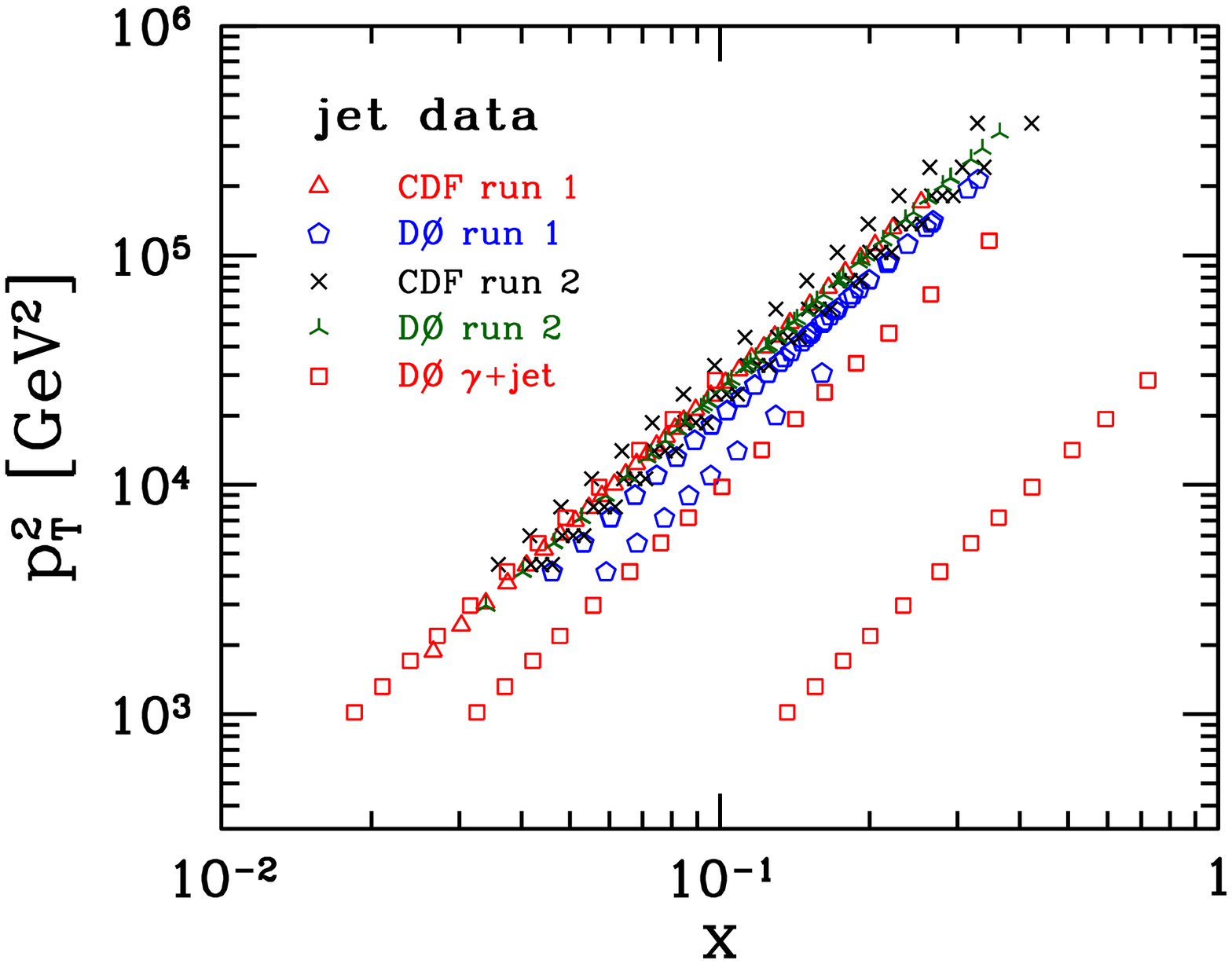}
\includegraphics[width=0.49\linewidth,trim= 0 120 0 10,clip=true]
                {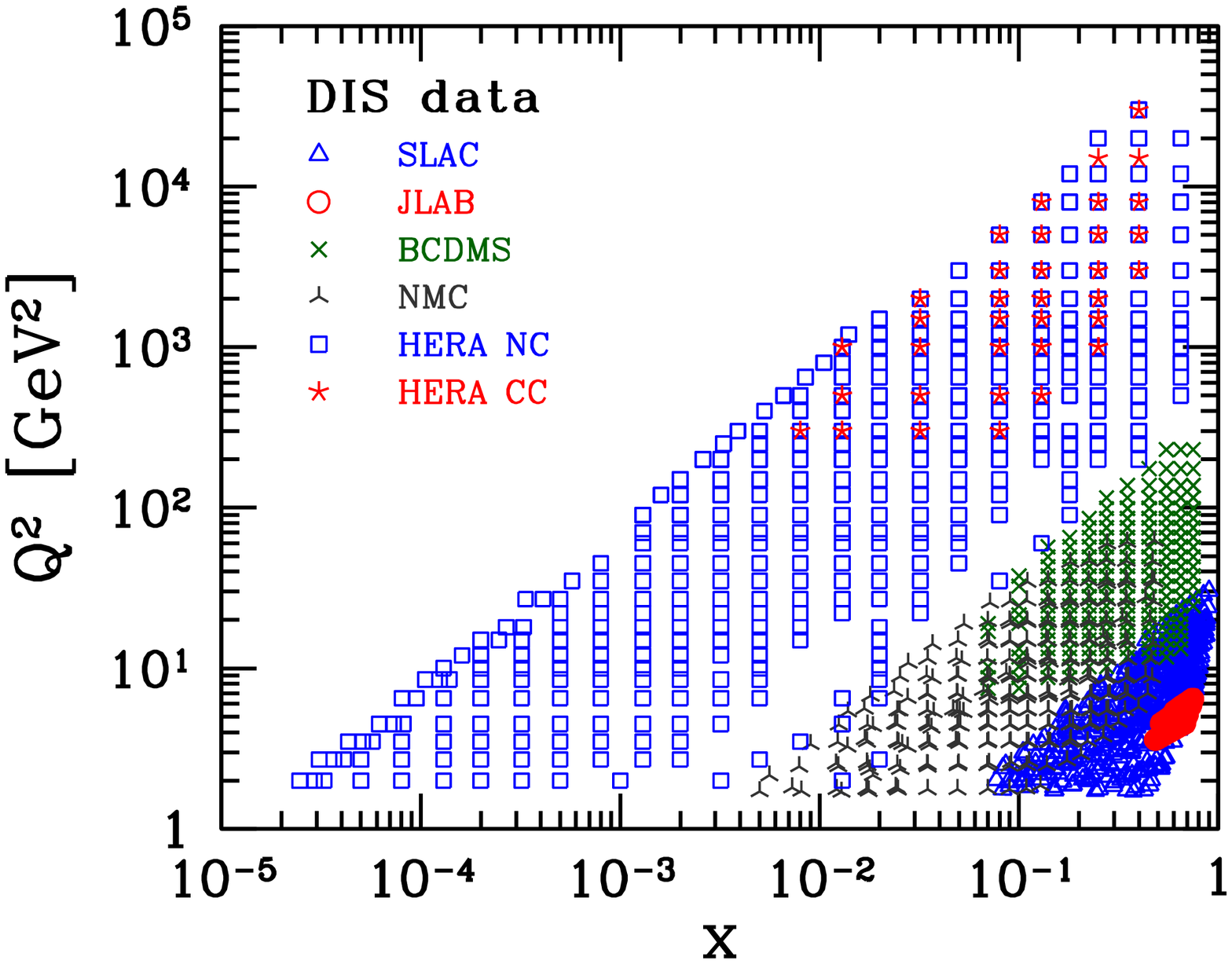}
\includegraphics[width=0.49\linewidth,trim= 0 120 0 10,clip=true]
                {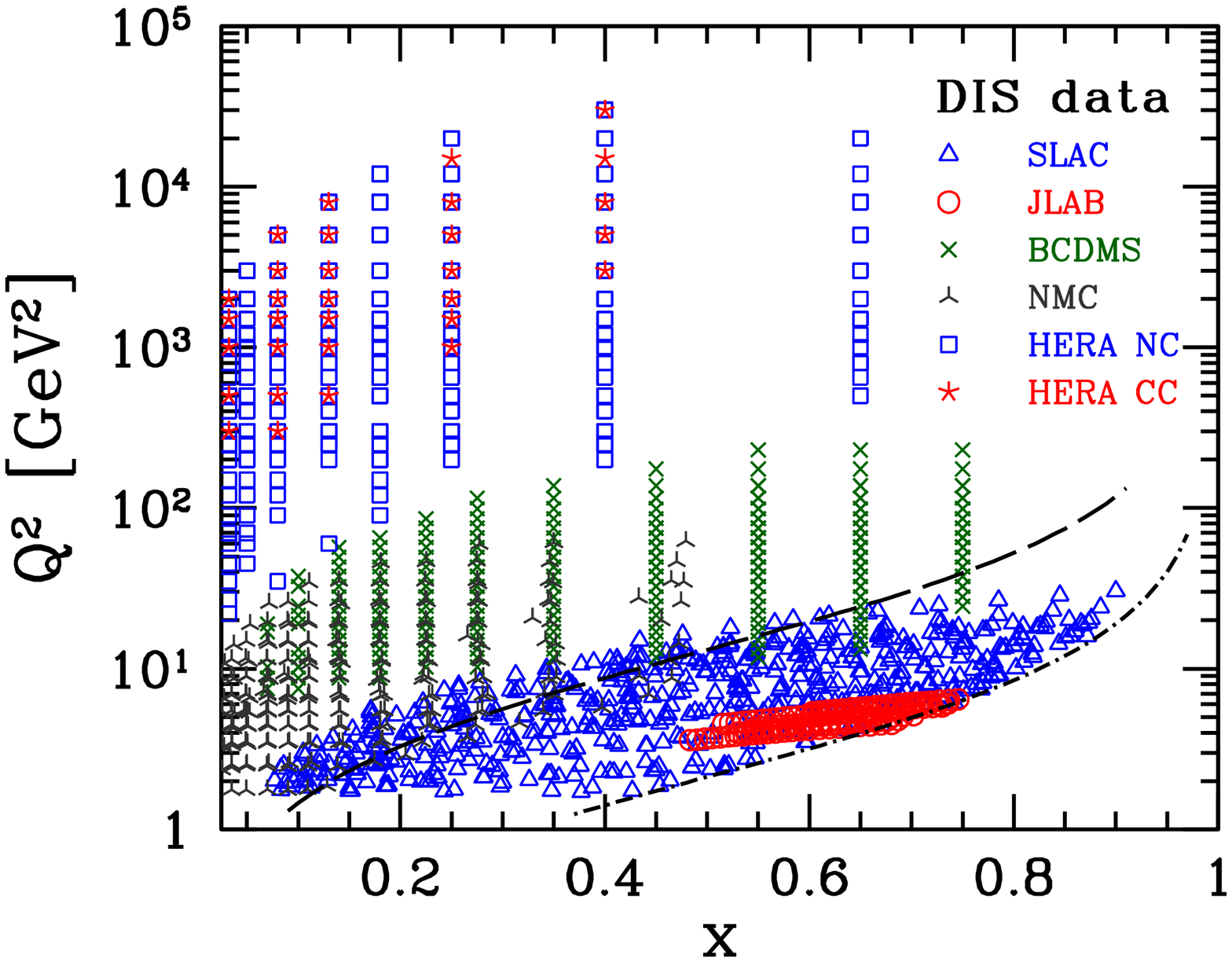}
\caption{Nominal coverage of the data sets used in the CJ12 global fits.
	{\it Top row}: vector boson and jet production.
	In the vector boson data plot, $M_V$ is the mass of the boson,
	and the weak boson data have been offset vertically for clarity.
	For the E866 dimuon data, the 3-pointed star and open triangle
	symbols correspond to the target and projectile $x$ values,
	respectively.
	In the jet data plot, $p_T$ is the transverse momentum of the jet.
	{\it Bottom row}: DIS data on a logarithmic and linear $x$ scale.
	In the latter, the $W^2 \gtrsim 14$~GeV$^2$ cut typically used
	in global PDF fits and the $W^2>3$ GeV$^2$ cut used in this
	analysis are indicated by the dashed and dot-dashed lines,
	respectively.}
\label{fig:xQ2coverage}
\end{figure}

The $x$ and $Q^2$ coverage of these data is illustrated in
Fig.~\ref{fig:xQ2coverage}, where leading order kinematics was
used according to the procedure detailed in Ref.~\cite{NNPDF10}.
While nominally there is a substantial amount of data at large $x$
from vector boson production measurements, only the directly
reconstructed $W^\pm$ charge asymmetry data from the CDF collaboration
has a sufficient statistical precision to constrain the $d$ quark
distribution uncertainty to the level aimed for in this study.
Note that the lepton charge asymmetry data also have high precision;
however, the lepton decay smears the primary $W$ production process,
resulting in an effective large-$x$ reach smaller than that for the
$W$ asymmetry.
The remaining source of large-$x$ data is DIS, for which the kinematic
cuts used in this work are comparable to those in the global fits of
Alekhin {\it et al.} \cite{ABKM09, ABM11}, but lower than the typical
cuts used in all other other global PDF analyses \cite{MSTW08, CT10,
JR09, HERAPDF10, NNPDF13}.  The improvement in the large-$x$ kinematic
reach is significant, amounting to about 1300 more data points roughly
equally divided between proton and deuteron targets, with adequate
precision.  This represents some $\sim 50\%$ increase in the total
number of DIS data points compared with the standard cuts.

\subsection{PDF parametrizations and conventions}

The calculations used in this analysis are all performed at NLO.
Correspondingly, the PDFs are evolved from the input scale $Q_0^2$
using NLO splitting functions.  It would be possible to utilize
NNLO splitting functions and, indeed, for some of the processes
the hard scattering cross section expressions are known to NNLO.
However, this is not true for the jet and $\gamma$ + jet processes.
In order to not mix NLO and NNLO treatments of data sets, we will
for consistency treat all fitted process at NLO.

For the $Q^2$ evolution, we use a zero-mass variable flavor scheme
with the charm and bottom flavor thresholds set at $m_c=1.3$~GeV
and $m_b=4.5$~GeV, respectively.  In this scheme, the only explicit
dependence on the quark masses is the value at which the number of
active flavors changes.  While it is true that various schemes exist
for taking into account the kinematic effects of nonzero quark masses
in DIS \cite{JR09}, those schemes have not been extended to the
other processes that have been included in this analysis.
Accordingly, we choose the zero-mass scheme so that the PDFs can
be treated consistently for each observable.

The fits described herein were all performed using a value of
$\Lambda^{(5)} = 0.2268$~GeV for 5 active flavors, corresponding
to the value $\alpha_s(M_Z) = 0.1180$ at two-loop accuracy.
This choice is based on the world average of
	$\alpha_s = 0.1184 \pm 0.0007$ \cite{PDG}.
Of course, one can also treat $\Lambda$ as a free parameter;
doing so, we find that the fitted $\Lambda$ results in a value
$\Lambda^{(5)} = 0.2278 \pm 0.0016$~GeV, consistent with our
fixed value.
However, the present analysis focusses on effects on the PDFs due
to the inclusion of the nuclear corrections for deuterium data.
Accordingly, we chose a fixed value of $\alpha_s(M_Z)$ specifically
to avoid having the coupling determined by one or more of the
higher statistics data sets.  We intend to return to this issue in
a future analysis by scanning a range of $\alpha_s(M_Z)$ values in
order to understand the effects on individual data sets.

For the parametrization of the PDFs at the input scale $Q_0^2$,
a common form has been adopted for all parton species $f$,
\begin{equation}
xf(x,Q_0^2) = a_0 x^{a_1} (1-x)^{a_2} (1 + a_3 \sqrt{x} + a_4 x).
\label{eq:param}
\end{equation}   
This form applies to the valence distributions
  $xq_v \equiv x(q-\bar q)$, for $q=u$ and $d$,
the isoscalar and isovector sea quark distributions
  $x(\bar u + \bar d)$ and $x(\bar d - \bar u)$,
and the gluon distribution $xg$.
However, to allow for a more flexible parametrization of the valence
$d_v$ PDF in the large-$x$ region, we add in a small admixture of the
$u_v$ PDF,
\begin{equation}
d_v \rightarrow
    a_0^{d_v} \left( \frac{d_v}{a_0^{d_v}} + b\, x^c u_v \right),
\label{eq:du}
\end{equation}
with $b$ and $c$ as two additional parameters.
The result of this modification is that
	$d_v/u_v \to a_0^{d_v}\, b$ as $x \to 1$,
provided $a_2^{d_v} > a_2^{u_v}$, which is usually the case.
A finite, nonzero value of this ratio is indeed expected in several
nonperturbative models of hadron structure \cite{FJ75, MT96, HR10}.
It is also required from a purely practical point of view, as it avoids
potentially large biases on the $d$-quark PDF central value \cite{CJ11},
as well as on its PDF error estimate, as we discuss in detail in
Sec.~\ref{sec:results}.
The $a_0$ parameters for the $u_v$ and $d_v$ distributions are fixed
by the appropriate valence quark number sum rules, while $a_0^g$ is
fixed by the momentum sum rule.
Finally, since we do not use neutrino--nucleus DIS data in the fits,
the strange quark PDF is assumed to be proportional to the isoscalar
light quark sea, and parametrized as
\begin{equation}
(s+\bar s) = \kappa (\bar u + \bar d).
\label{eq:kappa}
\end{equation}
The possibility of asymmetric $s$ and $\bar s$ distributions,
predicted in several nonperturbative models \cite{Sig87, MM96},
can be investigated using dimuon data in neutrino and antineutrino
DIS \cite{Mason07}, and will be the subject of future analysis
including nuclear corrections for heavy nuclei.
The parameter values for each of the three PDF sets (CJ12min,
CJ12mid and CJ12max) are given in Table~\ref{tab:parameters}
in Appendix~\ref{app:param}.

We find that the parametrization (\ref{eq:param}) is sufficiently
flexible to allow good fits to the combined data sets, with typical
$\chi^2$ values on the order of one per degree of freedom, as shown
in Table~\ref{tab:datasets}.  In some instances the parameters were
found to have very large errors, indicating that the chosen data sets
were unable to provide sufficiently strong constraints in those cases.
Such parameters were fixed at suitable values, and included in the
fits at the values shown in Table~\ref{tab:parameters} without errors.

\subsection{Target mass and higher twist corrections}

The standard method to calculate target mass corrections (TMCs)
in DIS is the one based on the operator product expansion in QCD,
first formulated by Georgi and Politzer \cite{GP76}.
An alternative method based on collinear factorization in momentum
space was developed by Ellis, Furmanski and Petronzio \cite{EFP},
and extended by various authors \cite{Kre04, AQ08, AHM09}
(see also the recent reviews of TMCs in Refs.~\cite{Sch08, Bra11}).
All of these methods suffer to some extent from the threshold problem,
whereby the target mass corrected structure function remains nonzero
as $x \to 1$ \cite{TungTMC, Ste06, Ste12}.
For the purposes of global fits, however, the region where the
threshold effects become problematic is $W \lesssim 2$~GeV \cite{Ste12},
which is mostly outside of where even the most liberal cuts in $W$ and
$Q^2$ are made \cite{CJ10, CJ11, ABKM09}.

According to the classic works of De~R\'ujula {\it et al.}
\cite{DGP77, DGPannals}, the appearance of the threshold problem is
attributed to the neglect of dynamical higher twist (HT) corrections,
which scale as powers in $1/Q^2$.  Moreover, higher order perturbative
QCD corrections can also resemble power suppressed contributions at
low $Q^2$.  Regardless of their origin, we can account for the various
power suppressed corrections that are not included in a leading twist
calculation by parametrizing them using a phenomenological multiplicative
factor, as in the earlier CJ fits \cite{CJ10, CJ11},
\begin{align}
F_2(x,Q^2)
= F_2^{\rm LT}(x,Q^2)
  \left( 1 + \frac{C(x)}{Q^2} \right),
\end{align}
where $F_2^{\rm LT}$ denotes the leading twist structure function
including TMCs.  For simplicity we generically refer to the fitted
$1/Q^2$ term as a ``higher twist'' correction, and parametrize the
higher twist coefficient function by
$C(x) = a_{\rm {\scriptscriptstyle HT}}\,
	x^{b_{\rm {\scriptscriptstyle HT}}}
	(1+c_{\rm {\scriptscriptstyle HT}} x)$,
assuming it to be isospin independent
(see, however, Refs.~\cite{Vir92, AKL03, BB08, Blu12}).

With the inclusion of TMCs and a phenomenological HT correction,
it was found in Ref.~\cite{CJ10} that the leading twist PDFs are
essentially independent of the TMC prescription adopted; the HT
parameters were able to compensate for the variations due to the
different TMC algorithms. 
While the focus in the present work is on the leading twist PDFs, the
interpretation and isospin dependence of the higher twist corrections
are of intrinsic interest in their own right, and will be the subject
of a future dedicated analysis.

\subsection{Nuclear corrections}
\label{sec:methods_nuclear}

Since nucleons bound in a nucleus are not free, the nuclear structure
function deviates from a simple sum of free proton and neutron
structure functions, especially at large $x$ where the effects of
Fermi motion, nuclear binding, and nucleon off-shellness are most
prominent.  In the nuclear impulse approximation the structure function
of the deuteron $d$ can be expressed as a convolution of the bound
nucleon structure function and a momentum distribution $f_{N/d}$ of
nucleons in the deuteron \cite{MSToff, KPW94, KP06},
\begin{align}
F_2^d(x,Q^2)\
=\ \sum_{N=p,n}
   \int dz\, f_{N/d}(z,\gamma)\, F_2^N (x/z,Q^2)\
+\ \delta^{(\rm off)} F_2^d(x,Q^2),
\end{align}
where the additive term $\delta^{(\rm off)} F_2^d(x,Q^2)$ represents
nucleon off-shell and relativistic corrections that cannot be expressed
through a one-dimensional convolution.
The momentum distribution, or ``smearing function'', $f_{N/d}$ is
computed in the weak binding approximation (WBA), which is appropriate
for a weakly bound nucleus such as the deuteron, in terms of the
deuteron wave function, and implements nuclear binding and Fermi
motion effects \cite{KP06, KMK09}.  At $Q^2 \to \infty$ it has a
simple probabilistic interpretation in terms of the light-cone momentum
fraction 
\mbox{$z = (M_d/M)(p \cdot q / p_d \cdot q)
   \approx (M_d/M)(p^+/p_d^+)$}
of the deuteron carried by the struck nucleon, where $p$ and $p_d$
are the four-momenta of the nucleon and deuteron, respectively,
and $M_d$ is the deuteron mass.
At finite $Q^2$, however, the smearing function depends in addition
on the parameter \mbox{$\gamma^2 = 1 + 4x^2 M^2/Q^2$}, which
characterizes the deviation from the Bjorken limit.

The nuclear corrections at large $x$ depend partly on the strength
of the high-momentum tail of the deuteron wave function, and we use
several wave functions based on different nucleon--nucleon potentials
to study the deuteron model dependence.  We choose the high-precision
AV18 \cite{AV18}, CD-Bonn \cite{CDBonn} and the relativistic WJC-1
\cite{WJC} wave functions, which provide a representative spread of
behaviors at high momentum.
Note that the effects of the nuclear smearing corrections are not
suppressed at large $Q^2$, and must be considered at all scales
wherever data at $x \gtrsim 0.5$ are used \cite{CJ10, ACHL09, ARM12}.

The off-shell nucleon correction $\delta^{(\rm off)} F_2^d$ is somewhat
more model dependent, but several quark model based estimates of this
have been made in the literature \cite{KP06, GL92, MSTplb, MSS97}.
In this work we follow the approach adopted in our previous analysis
\cite{CJ11}, which utilized the ``modified Kulagin-Petti'' model. 
In this model, the corrections were related to the change in the
nucleon's confinement radius in the nuclear medium, as well as the
average virtuality of the bound nucleons, and constrained to give no
net change in the structure function normalization.  In contrast to
Ref.~\cite{CJ11}, however, here we further take into account the
correlation between the nucleon ``swelling'' and the deuteron wave
function.  The combined effects introduce a theoretical uncertainty
in the extracted PDFs, particularly for the $d$ quark.  A detailed
evaluation of the impact of this uncertainty will be presented in
Sec.~\ref{sec:results}.

\subsection{PDF Errors}
\label{sec:errors}

The global fits were performed using a standard $\chi^2$ minimization
algorithm due to Marquardt \cite{Marquardt}, as described in
Ref.~\cite{Bevington}.  This prescription produces a well-defined
Hessian matrix typically in fewer than a dozen iterations and is
both efficient and well-behaved.  Where available, correlated
systematic errors have been utilized.  Error PDF sets were produced
using the resulting Hessian following the methods described in
Ref.~\cite{Cteq6}.  Two error PDFs were produced for each of the
eigenvalues of the Hessian, corresponding to positive and negative
parameter increments, resulting in a total of 38 error PDFs for each
of the three choices of nuclear corrections.

Using these error PDFs it is possible to construct an error estimate
$\delta \sigma$ for some physical observable $\sigma$ via 
\begin{equation}
\delta \sigma
= \frac{T}{2}
  \sqrt{\sum_{i=1}^{19}
    \Big[ \sigma(a_{2i-1}) - \sigma(a_{2i}) \Big]^2},
\label{eq:sym_er}
\end{equation}
where $a_i$ denotes the set of parameters corresponding to the PDF
set $i$, and $T$ is a $\chi^2$ tolerance that will be described below.
Alternatively, one can define asymmetric upper and lower errors
$\delta \sigma_+$ and $\delta \sigma_-$ based on the same error PDF
sets using the expressions
\begin{subequations}
\begin{align}
\delta \sigma_+
& = T \sqrt{ \sum_{i=1}^{19}
      \Big( \max\Big[ \sigma(a_{2i-1})-\sigma(a),
		      \sigma(a_{2i})-\sigma(a),
			0
		\Big]
      \Big)^2 },
\label{eq:asym1}			\\
\delta \sigma_-
& = T \sqrt{\sum_{i=1}^{19}
      \Big( \max\Big[ \sigma(a)-\sigma(a_{2i-1}),
		      \sigma(a)-\sigma(a_{2i}),
			0
		\Big]
      \Big)^2 },
\label{eq:asym2}
\end{align}
\end{subequations}
where $\sigma(a)$ denotes the observable calculated with the 
``central value'' of the fitted PDF set.

The issue of the appropriate value for the $\chi^2$ tolerance $T$
has been discussed at length in the literature.  The value $T=1$
is suitable in the case where all data sets are compatible with
each other and all errors are Gaussian.
In practice, various tolerance values are used in different PDF
analyses, including $T=10$ for the CTEQ6 PDFs \cite{Cteq6},
$T=\sqrt{50}$ in the MRST2002 fits \cite{MRST02}, and dynamically
determined values in more recent fits \cite{MSTW08, CT10}.
In the CJ12 analysis discussed in this paper the error bands have
been calculated using $T=10$.

\subsection{Theoretical errors}

A significant source of theoretical errors is the modeling of nuclear
corrections for DIS on deuterium targets, which we refer to as
``nuclear uncertainty'' in short.  Of course, since the deuteron
can never be considered as composed of free nucleons, it is never
physically reasonable to assume no nuclear corrections at all.
We evaluate the nuclear uncertainty by considering nuclear corrections
ranging from mild, corresponding to the hardest of the modern deuteron
wave functions (WJC-1 \cite{WJC}) coupled to a 0.3\% nucleon swelling
effect, to strong, corresponding to the softest of the wave functions
(CD-Bonn \cite{CDBonn}) and a large, 2.1\% nucleon swelling.
An intermediate nuclear correction corresponding to the AV18
\cite{AV18} deuteron wave function and a 1.2\% nucleon swelling
effect is taken for our central fit.  These corrections approximately
span the same range as in our previous analysis \cite{CJ11}.

\begin{figure}
\includegraphics[scale=0.4]{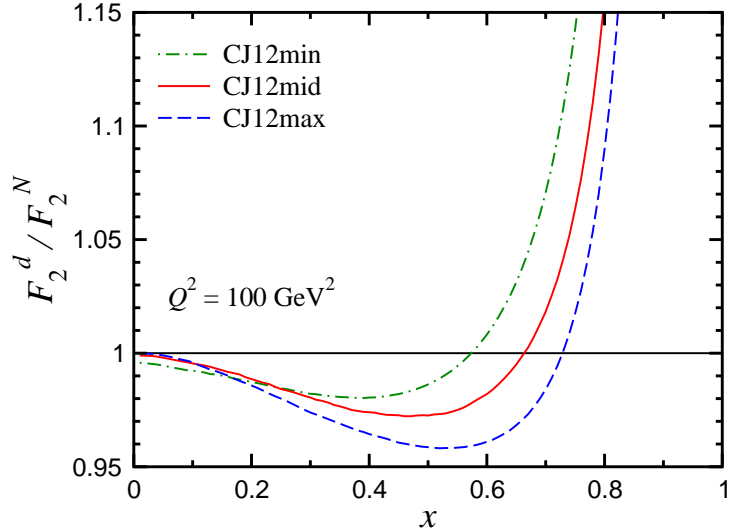}
\caption{Ratio of deuteron to isoscalar nucleon structure functions
	$F_2^d/F_2^N$ for the three sets of nuclear corrections
	considered in this analysis, corresponding to the
	CJ12min (dot-dashed), CJ12mid (solid), and CJ12max (dashed)
	PDFs.}
\label{fig:d2N}
\end{figure}

The effect of the different nuclear corrections is illustrated in
Fig.~\ref{fig:d2N} for the ratio of $F_2$ structure functions of the
deuteron and an idealized isoscalar target consisting of an unbound
proton and neutron.  The upper, middle, and lower curves are the
results for the mild (CJ12min), intermediate (CJ12mid), and strong
(CJ12max) nuclear corrections.  An intuitive way to understand the
effects of these nuclear corrections on global fits is to consider the
idealized free proton plus neutron target as being obtained by dividing
the deuterium target data by the ratio shown in Fig.~\ref{fig:d2N}.
In the intermediate-$x$ region the idealized nucleon $F_2^N$ structure
function is larger than the original data, and the increase is largest
for the maximum nuclear correction model.  On the other hand, as one
extends to values of $x$ above about 0.75, the reverse is true: the
$F_2^N$ results there lie below the data and are lowest for the mild
set of corrections.

In a global fit, it is the combination of PDFs and nuclear corrections
which is constrained by deuterium data, not the PDFs alone.
At large $x$, the $u$ quark is well constrained by free proton DIS
and other data, so that when fitting the corrected deuterium data
(essentially the idealized $F_2^N \propto u+d$) the $d$ quark will
be the distribution most sensitive to the nuclear corrections.
Therefore, in the large-$x$ region the $d$ PDF will be smallest
for the mild correction and largest for the strong corrections
(see Fig.~\ref{fig:du} in Sec.~\ref{sec:results}).

One way to evaluate the impact of the nuclear uncertainty on a given
observable is by calculating it using the CJ12min fit corresponding
to the smallest nuclear correction, then with the CJ12max fit
corresponding to the largest correction. The values of the observable
spanning these 2 extremes can be considered a fair representation of
the theoretical nuclear uncertainty.  PDF errors can then be added
to the central values so calculated, and displayed as an outer error
band, as done in Ref.~\cite{CJ10, CJ11}.  Alternatively, one could first
calculate a PDF error band using the CJ12mid set, and display this on
top of the 2 bands obtained with the min and max sets.  The size of
the resulting sidebands will then represent the nuclear uncertainty.

\section{Results} 
\label{sec:results}

In this section the results of the CJ12min, CJ12mid, and CJ12max
sets of global PDF fits are presented and compared with other modern
PDF fits.  Tables for the central PDF sets, the corresponding
error PDF sets, and a user interface to read them can be found
on the CJ \cite{CJweb} and CTEQ \cite{CTEQweb} web sites.

The data sets used in the global fits are shown in
Table~\ref{tab:datasets} along with the $\chi^2$ values of
the three fits for each of the data sets used in the analysis,
as well as the total $\chi^2$ values and the totals with the
normalization contribution to the $\chi^2$ included.
As is apparent from these values, the description of most of the
data sets is independent of the size of the nuclear corrections.
Indeed, the deuterium data are sensitive to the convolution of
the nuclear model with the $d$ PDF; as the model is changed,
the $d$ PDF is altered by the fitting program to compensate.
For observables such as the jet cross section, the modified $d$
PDF can be compensated for by small changes in the gluon PDF,
so the overall description of the data remains unaltered.
One notable exception to this pattern is for the $W$ asymmetry
as measured by the CDF collaboration \cite{CDF09}.
This observable is sensitive to the $d/u$ ratio, and changes
in the $d$ PDF are not easily compensated by the $u$ PDF,
which is relatively well constrained by proton DIS data.
The $\chi^2$ for the $W$ asymmetry data does show a significant
increase as the magnitude of the nuclear corrections increases
beyond its middle value, indicating a preference for mild to
medium nuclear corrections.

In this regard it is interesting to compare our results to those of
the recent analysis in Ref.~\cite{MMSTWW}, which included nuclear
corrections for deuterium targets in DIS using a 4-parameter,
$Q^2$-independent phenomenological function with the parameters
varied in the fit.  The resulting correction factor, shown in Fig.~11
of Ref.~\cite{MMSTWW}, can be compared to those in Fig.~\ref{fig:d2N}
above.  Their fitted form lies between the curves for the CJ12min and
CJ12mid fits, as might be expected since these two fits have nearly
identical values for $\chi^2$, while the CJ12max value is higher.
As noted above, much of the increase in $\chi^2$ for the CJ12max set
is due to the CDF $W$ asymmetry data, which is also included in the
fit of Ref.~\cite{MMSTWW}.  Although this comparison is not exact,
since our nuclear corrections are $Q^2$ dependent \cite{Acc11}
and those in Ref.~\cite{MMSTWW} are not, it is consistent with our
observation that the nuclear model choices made for the CJ12min and
CJ12mid sets are preferred by the data.

\begin{figure}
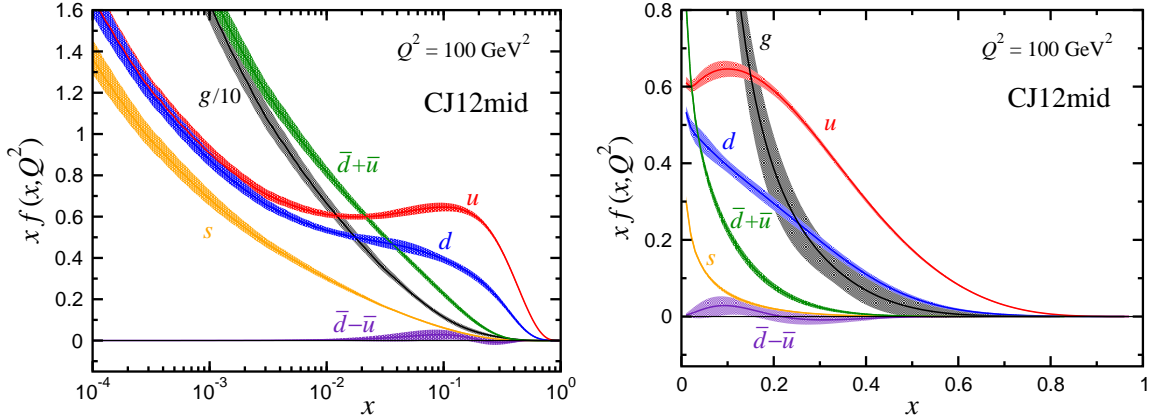

\includegraphics[height=5.5cm]{Fx_errl.eps}\ \
\includegraphics[height=5.5cm]{Fx_err.eps}
\caption{Uncertainty bands for the $u$, $d$, $\bar d+\bar u$,
	$\bar d-\bar u$, $s$ and $g$ PDFs for the CJ12mid fit
	at $Q^2=100$~GeV$^2$, shown on logarithmic (left) and
	linear (right) scales in $x$.  Note that in the left
	panel the gluon is scaled by 1/10.}
\label{fig:CJ12}
\end{figure}

\begin{figure}
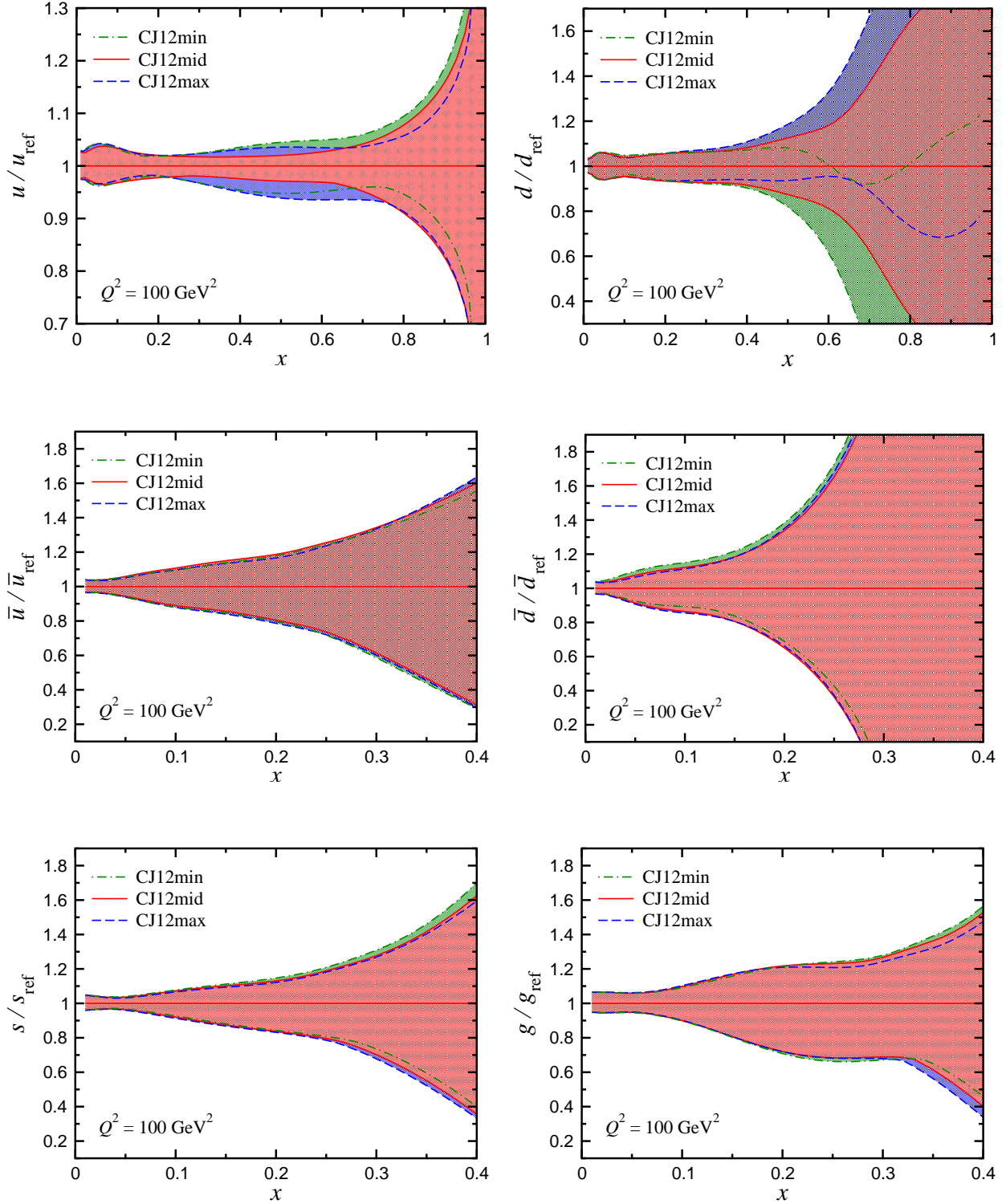

\includegraphics[width=8cm]{relerrCJu.eps}\ \ \
\includegraphics[width=8cm]{relerrCJd.eps}\vspace*{1cm}
\includegraphics[width=8cm]{relerrCJub.eps}\ \ \
\includegraphics[width=8cm]{relerrCJdb.eps}\vspace*{1cm}
\includegraphics[width=8cm]{relerrCJs.eps}\ \ \
\includegraphics[width=8cm]{relerrCJg.eps}
\caption{PDF uncertainties for the CJ12min (dot-dashed), CJ12mid (solid),
	and CJ12max (dashed) PDF sets relative to the reference CJ12mid
	set at $Q^2=100$~GeV$^2$.}
\label{fig:relCJ12}
\end{figure}

The CJ12mid PDFs are shown in Fig.~\ref{fig:CJ12} at $Q^2=100$~GeV$^2$
with the PDF error bands calculated as described in
Sec.~\ref{sec:errors}, on both logarithmic and linear $x$ scales.
The latter more graphically illustrates the behavior of the PDFs
at large values of $x$, where the uncertainties from nuclear and
finite-$Q^2$ corrections are greatest.  The error bands are shown
in more detail in Fig.~\ref{fig:relCJ12}, and compared to the
CJ12min and CJ12max sets.  It is clear that the effects of nuclear
corrections are strongest on the $d$ PDF, with the others showing
little or no influence.  In particular, the $d$ quark PDF increases
at large values of $x$ as the magnitude of the nuclear corrections
increases.  This, again, is a reflection of the fact that it is the
convolution of the nuclear corrections and the $d$ PDF that is
constrained by the deuterium DIS data, and not the $d$ PDF alone.

\begin{figure}
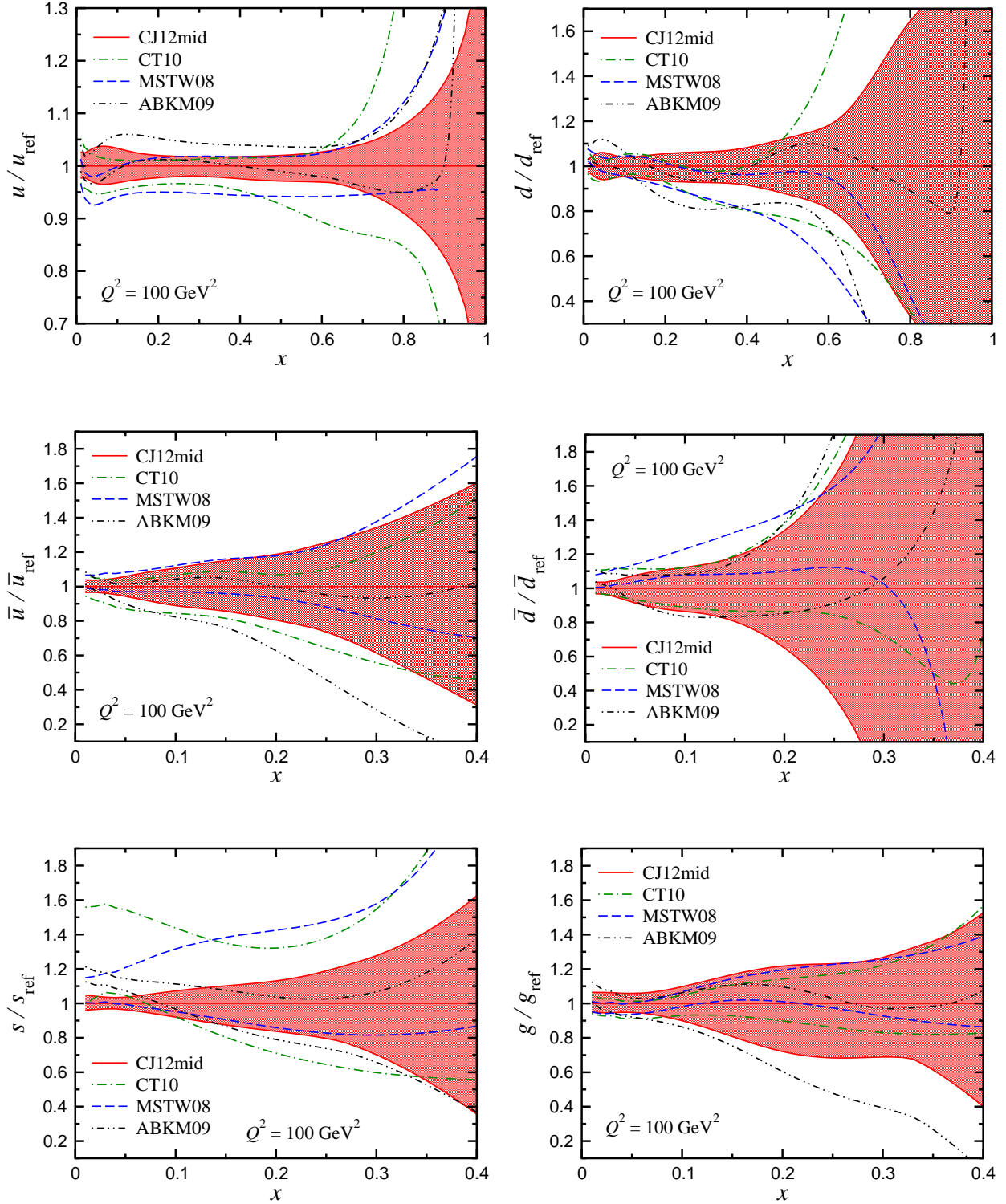

\includegraphics[width=8cm]{relerrPDFu.eps}\ \ \
\includegraphics[width=8cm]{relerrPDFd.eps}\vspace*{1cm}
\includegraphics[width=8cm]{relerrPDFub.eps}\ \ \
\includegraphics[width=8cm]{relerrPDFdb.eps}\vspace*{1cm}
\includegraphics[width=8cm]{relerrPDFs.eps}\ \ \
\includegraphics[width=8cm]{relerrPDFg.eps}
\caption{PDF uncertainties for the CJ12mid (solid), CT10 (dot-dashed),
	MSTW08 (dashed) and ABKM09 (dot-dot-dashed) PDF sets
        relative to the reference CJ12mid set at $Q^2=100$~GeV$^2$.}
\label{fig:relPDF12}
\end{figure}

A comparison of the CJ12mid PDFs with several other sets is shown
in Fig.~\ref{fig:relPDF12}, which for illustration includes the
CT10 \cite{CT10}, MSTW08 \cite{MSTW08} and ABKM09 \cite{ABKM09} PDFs.
The latter in particular are closest in spirit to the CJ12 PDFs in
that similar kinematic cuts are imposed, and nuclear and finite-$Q^2$
corrections included, albeit using somewhat different prescriptions.
One feature that is readily apparent in the first two panels is that
the error bands for the $u$ and $d$ PDFs are reduced relative to those
from the other sets.  This is a direct consequence of the lower cuts
on $Q^2$ and $W^2$ which allow DIS data to be fitted to larger values
of $x$, approaching 0.9.  Of course, when one considers the variations
allowed by the different nuclear models, the error bands again increase,
modestly for the $u$ quark and substantially for the $d$ quark.

\begin{figure}
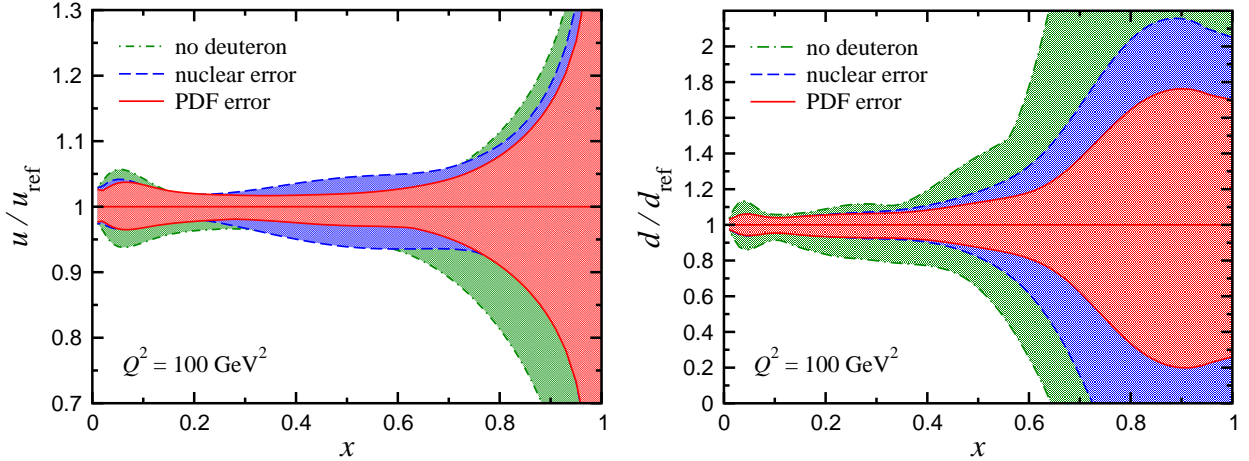

\includegraphics[width=8cm]{relerrNODu.eps}\ \ \ 
\includegraphics[width=8cm]{relerrNODd.eps}
\caption{PDF uncertainties for the CJ12mid $u$ and $d$ quark PDFs
	(solid) compared with the total uncertainty from nuclear
	corrections (dashed) and with a fit excluding all deuterium
	data (dot-dashed), relative to the reference CJ12mid set at
	$Q^2=100$~GeV$^2$.}
\label{fig:nod}
\end{figure}

To further examine the role of the deuteron DIS data, a global fit
was performed with the deuterium DIS points removed from the data set.
The results are shown in Fig.~\ref{fig:nod} for the $u$ and $d$ PDFs
relative to those from the CJ12mid set.  Without the deuterium data
the error band for the $u$ PDF shows a modest increase beyond
$x \approx 0.7$, while that for the $d$ PDF shows a significant
increase over most of the $x$ range.  It is important to note that
the error band on the $d$ PDF without the deuterium data exceeds
the band which includes the nuclear uncertainties.  This clearly
demonstrates the usefulness of the deuterium DIS data, even in the
presence of the nuclear uncertainties that its use introduces.

For the strange quark PDF, the error band shown is somewhat smaller
than for other sets.  This is an artificial reduction, however, since
we do not use the NuTeV dimuon DIS data \cite{Mason07}, which are
taken on heavy nuclear targets.  Accordingly, the value of $\kappa$
in Eq.~(\ref{eq:kappa}) is fixed, rather than fitted, and the error
on the $s+\bar s$ combination is therefore proportional to that of
$\bar u + \bar d$.  In practice, we use the value $\kappa=0.4$ from
earlier CTEQ fits \cite{Cteq6, Cteq61} that employed the same choice
for $Q_0$ as in this analysis.  Methods of constraining the $s$ PDF
without the use of heavy nuclear targets will be the subject of
future investigation.

\begin{figure}
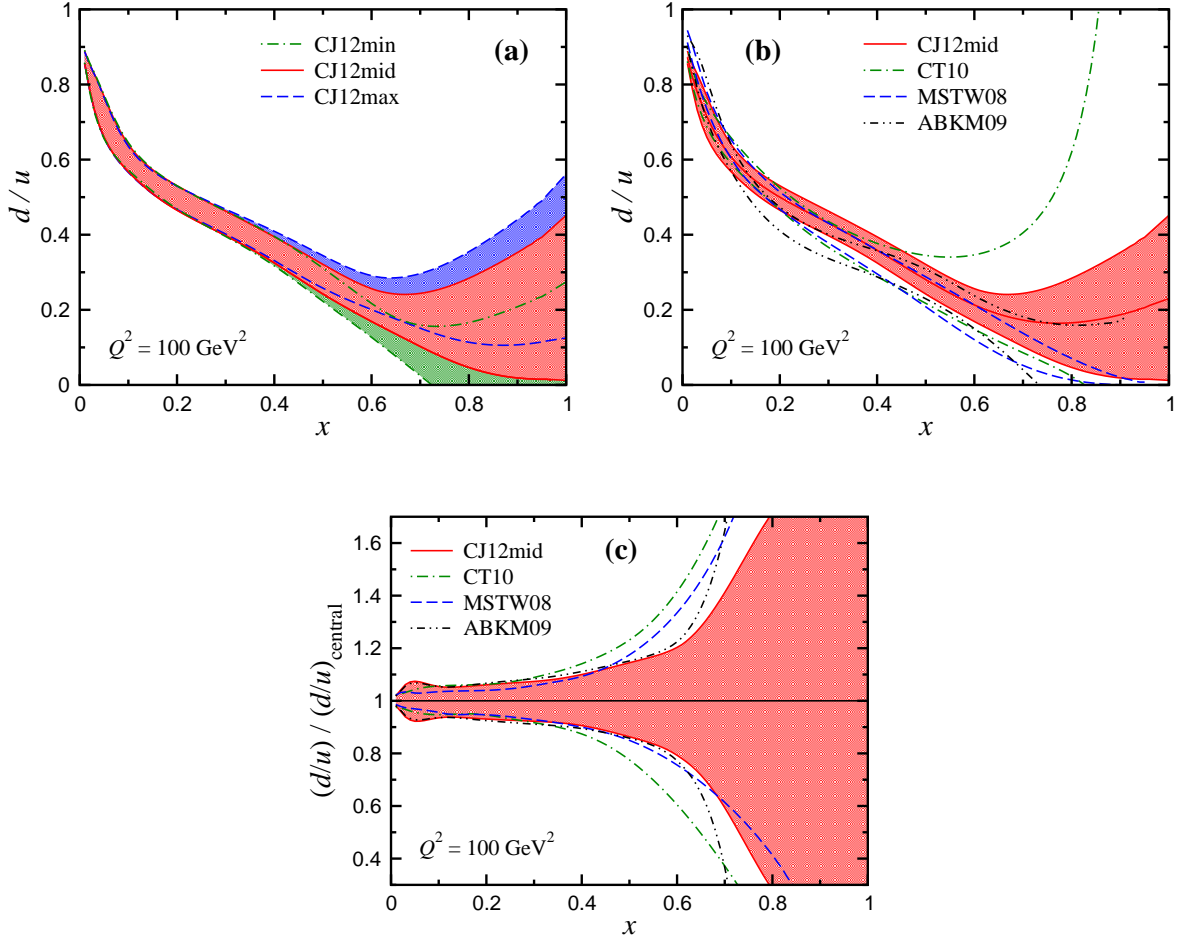

\includegraphics[width=7.5cm]{relerrCJdu.eps}\ \ \ \
\includegraphics[width=7.5cm]{relerrPDFdu.eps}\vspace*{1cm}
\includegraphics[width=7.5cm]{relerrPDFdu_rel.eps}
\caption{$d/u$ ratio at $Q^2=100$~GeV$^2$ for
	(a) the CJ12min (dot-dashed), CJ12mid (solid) and
	    CJ12max (dashed) PDFs,
        (b) the CJ12mid (solid), CT10 (dot-dashed),
	    MSTW08 (dashed) and ABKM09 (dot-dot-dashed) PDFs, and
	(c) relative to the central values of $d/u$ for
	    each PDF set.}
\label{fig:du}
\end{figure}

The ratios of the $d$ to $u$ PDFs for the three CJ12 sets are shown
in Fig.~\ref{fig:du}, which also compares the CJ12mid results with
those from the other PDF sets in Fig.~\ref{fig:relPDF12}.
Conventional parametrizations that treat the $d$ and $u$ PDFs
with the same $(1-x)^{a_2}$ form for the $x \to 1$ behavior,
have the characteristic that the $d/u$ ratio tends either to
zero or infinity as $x$ approaches one.
Conversely, the use of the parametrization given in Eq.~(\ref{eq:du})
allows $d/u$ to have any finite limiting value as $x \to 1$.
This is clear from the results in Fig.~\ref{fig:du}(b), where the
MSTW08 and ABKM09 $d/u$ ratios tend to zero, while that for CT10
tends to infinity.  This behavior distorts the error bands so that
the MSTW08 ratio appears to have an anomalously small uncertainty,
while the CT10 result is much larger.  However, as the relative
errors in Fig.~\ref{fig:du}(c) show, the MSTW08 and CT10 bands
are similar, as would be expected, while the CJ12 and ABKM09 bands
are somewhat reduced in size at larger values of $x$ due to the
additional high-$x$ data used in those analyses.  The ABKM09 band
then diverges again from the CJ12 band because of the use of a
conventional $d$ quark parametrization.

As the magnitude of the nuclear corrections increases, the value of
the $d/u$ intercept at $x=1$ rises because of the mechanism discussed
in Sec.~\ref{sec:methods_nuclear}.  The central values of $d/u$
extrapolated to $x=1$ are 0.012, 0.22, and 0.33 for the CJ12min,
CJ12mid, and CJ12max fits, respectively.  Including the PDF errors
we find
\begin{align}
  d/u \,\xrightarrow[\,x\rightarrow 1\,]{} \, 0.22
	\pm 0.20 \, \text{({\small PDF})}
	\pm 0.10 \,\text{(nucl)},
\end{align}
where the first error is from the PDF fits and the second is from
the nuclear correction models.  These values encompass the range
of available theoretical predictions \cite{Feyn72, Close73, FJ75,
MT96, HR10}.  However, it is also clear that a relatively modest
improvement in statistical precision and reduction of nuclear
uncertainty would allow one to restrict the range of allowable
physical mechanisms.

\section{Conclusion}
\label{sec:conclusion}

The CJ12 next-to-leading order PDFs presented in this study
demonstrate the clear necessity of including nuclear corrections
in global fits for PDFs when deuterium DIS data are included.
The largest effect is on the $d$ quark PDF, where variations
can be observed for values of $x \gtrsim 0.5$.
The theoretical uncertainty accompanying these corrections is of the
same order of magnitude as the PDF uncertainty, and the CJ12 PDF sets
will enable one to assess the impact of nuclear effects on the error
bands estimated for various observables.  The impact of these
corrections is important for observables that depend on PDFs at
moderate to large values of $x$, such as the production of high
mass states and cross sections at large values of rapidity.
Examples of such effects for selected observables at the Tevatron
and LHC were recently discussed in Ref.~\cite{Bra12}.

The CJ12 fits utilize an expanded DIS data set obtained by relaxing
the kinematic cuts to $Q^2 > Q_0^2 = (1.3\ \rm GeV)^2$ and
$W^2 > 3$~GeV$^2$, and considering finite-$Q^2$ corrections
such as target mass and higher twist effects in the theoretical
calculations.  The resulting significant improvement in the statistics
of the DIS data at large $x$ more than compensates for the nuclear
correction uncertainty, resulting in a net improvement in the
precision to which the $d$ quark is extracted from experimental data.

The PDFs presented in this analysis form a baseline for comparisons 
to be done in further studies, and there are a number of issues to be investigated. For example, we would like to determine the effects on individual data sets of changing the $\alpha_s(M_Z)$ value used in the fit. Similarly, we envision a dedicated study of the effects of choosing different conventions for treating heavy quarks in DIS processes and of including nuclear corrections for other observables.

Further refinements of the CJ12 PDFs are anticipated as additional
data become available.  One natural extension is to assess the
constraints provided by the addition of new LHC data to the
global fits, such as on the $s$ quark distribution \cite{ATLAS-s}.
Additional constraints on other PDFs may be provided through $Z$
rapidity measurements \cite{LHCb}, and high-$p_T$ jet and direct
photon production \cite{directgamma}.
A dedicated study of the impact of longitudinal structure function
measurements at Jefferson Lab \cite{Lia04, Mon12} on large-$x$ gluons
is also foreseen.  The potential of large rapidity measurements in
proton-proton collisions at the upgraded RHIC detectors, or the
proposed AFTER@LHC experiment \cite{AFTER@LHC} can also be explored.

As new and more precise measurements of observables sensitive to the
$d$-quark in the proton and less sensitive to nuclear corrections
become available in the future \cite{MARATHON, BONUS12, SOLID},
complementing the existing $W$ charge asymmetry data \cite{CDF09},
global QCD fits will become capable of constraining theoretical models
of nuclear corrections in the deuteron.  Not only will this reduce the
nuclear uncertainty on the fitted PDFs, it will also provide a new
avenue for progress in the theoretical understanding of high-energy
processes involving nuclei.

\acknowledgments

We thank M.~E.~Christy, P.~Jimenez-Delgado, C.~E.~Keppel and P.~Monaghan
for helpful discussions, and M.~E.~Christy and S.~Malace for assistance
with the parametrizations of the nucleon off-shell corrections used here.
This work was supported by the DOE contract No.~DE-AC05-06OR23177,
under which Jefferson Science Associates, LLC operates Jefferson Lab.
The work of J.F.O. and A.A. was supported in part by DOE contracts
No.~DE-FG02-97ER41922 and No.~DE-SC0008791, respectively.

\newpage
\appendix
\section{Parameter values}
\label{app:param}

In this appendix we list the initial parameter values and their errors
for the three sets of PDFs discussed in the text.

\begin{table}[htb]
\begin{center}
\caption{Parameter values for the three CJ12 PDF sets at the initial
	scale $Q_0 = 1.3$~GeV.  The parameters without errors have
	been fixed by sum rules or other constraints.\\}
{\scriptsize
\begin{tabular}{c|c|c|c}\hline
Parameter & CJ12min & CJ12mid & CJ12max \\ \hline
$a_0^{u_v}$
	& 2.8468
	& 2.7684
	& 2.9343 			\\
$a_1^{u_v}$
	& 0.65468 $\pm$ 0.02138
	& 0.64839 $\pm$ 0.02115
	& 0.66220 $\pm$ 0.02152 	\\
$a_2^{u_v}$
	& 3.5453 $\pm$ 0.0178
	& 3.5686 $\pm$ 0.0164
	& 3.5733 $\pm$ 0.0175		\\
$a_3^{u_v}$
	& 0.0
	& 0.0
	& 0.0				\\
$a_4^{u_v}$
	& 2.7201 $\pm$ 0.4322
	& 2.8970 $\pm$ 0.4005
	& 2.6616 $\pm$ 0.4266		\\ \hline
$a_0^{d_v}$
	& 15.389
	& 21.515
	& 24.724 			\\
$a_1^{d_v}$
	& 1.0296 $\pm$ 0.0449
	& 1.1004 $\pm$ 0.0422
	& 1.1328 $\pm$ 0.0420		\\
$a_2^{d_v}$
	& 6.3767 $\pm$ 0.1860
	& 6.9525 $\pm$ 0.1905
	& 7.1420 $\pm$ 0.1947		\\
$a_3^{d_v}$
	& $-3.7930 \pm 0.1098$
	& $-3.9659 \pm 0.0984$
	& $-3.9989 \pm 0.0949$		\\
$a_4^{d_v}$
	& 5.8404 $\pm$ 0.2808
	& 5.9770 $\pm$ 0.2454
	& 5.9310 $\pm$ 0.2326 		\\
$b$
	&\ $(0.0787 \pm 0.1648)\times10^{-2}$\
	&\ $(1.0379 \pm 0.1688)\times10^{-2}$\
	&\ $(1.3464 \pm 0.2187)\times10^{-2}$\	\\
$c$
	& 2.0 & 2.0 & 2.0		\\ \hline
$a_0^{\bar u+\bar d}$
	& 0.12145 $\pm$ 0.00491
	& 0.12154 $\pm$ 0.00486
	& 0.12366 $\pm$ 0.00490		\\
$a_1^{\bar u+\bar d}$
	& $-0.23390 \pm 0.00445$
	& $-0.23389 \pm 0.00441$
	& $-0.23222 \pm 0.00438$	\\
$a_2^{\bar u+\bar d}$
	& 9.6939 $\pm$ 0.1901
	& 9.7127 $\pm$ 0.1917
	& 9.5842 $\pm$ 0.1975		\\
$a_3^{\bar u+\bar d}$
	& 0.0
	& 0.0
	& 0.0				\\
$a_4^{\bar u+\bar d}$
	& 23.689 $\pm$ 1.740
	& 23.228 $\pm$ 1.704
	& 21.882 $\pm$ 1.649		\\ \hline 
$a_0^{\bar d-\bar u}$
	&  90.16 $\pm$ 21.18
	& 112.02 $\pm$ 25.64
	& 115.65 $\pm$ 26.69 		\\
$a_1^{\bar d-\bar u}$
	& 2.5411 $\pm$ 0.1428
	& 2.7048 $\pm$ 0.1477
	& 2.7408 $\pm$ 0.1492		\\
$a_2^{\bar d-\bar u}$
	& $a_2^{\bar u+\bar d}$ + 2.5
	& $a_2^{\bar u+\bar d}$ + 2.5
	& $a_2^{\bar u+\bar d}$ + 2.5	\\
$a_3^{\bar d-\bar u}$
	& 0.0
	& 0.0
	& 0.0				\\
$a_4^{\bar d-\bar u}$
	& $-4.1657 \pm 0.1932$
	& $-4.2848 \pm 0.1953$
	& $-4.1870 \pm 0.1914$		\\ \hline
$a_0^g$
	& 46.911
	& 49.404
	& 51.157			\\
$a_1^g$
	& 0.61077 $\pm$ 0.03228
	& 0.62122 $\pm$ 0.03360
	& 0.63027 $\pm$ 0.03479	\\
$a_2^g$
	& 6.1691 $\pm$ 0.6198
	& 6.5242 $\pm$ 0.6703
	& 6.8007 $\pm$ 0.7273 		\\
$a_3^g$
	& $-3.5055 \pm 0.1251$
	& $-3.5348 \pm 0.1337$
	& $-3.5298 \pm 0.1423$		\\
$a_4^g$
	& 3.5292 $\pm$ 0.2777
	& 3.6436 $\pm$ 0.3059
	& 3.6777 $\pm$ 0.3324 		\\ \hline
$\kappa$
	& 0.4
	& 0.4
	& 0.4				\\ \hline
$a_{\rm {\scriptscriptstyle HT}}$
        & $-0.5627 \pm 1.3948$
        & $-0.5679 \pm 1.1627$
        & $-0.4992 \pm 1.2213$		\\
$b_{\rm {\scriptscriptstyle HT}}$
        & $3.3528 \pm 0.9203$
        & $3.1936 \pm 0.8423$
        & $3.1712 \pm 0.8654$		\\
$c_{\rm {\scriptscriptstyle HT}}$
        & $-9.042 \pm 20.930$
        & $-8.493 \pm 16.200$
        & $-9.459 \pm 21.695$		\\ \hline
\end{tabular}
}
\label{tab:parameters}
\end{center}
\end{table}

\newpage


\begin{thebibliography}{99}

\bibitem{Feyn72}  
R.~P.~Feynman, {\em Photon Hadron Interactions}
(Benjamin, Reading, Massachusetts, 1972).

\bibitem{Close73}
F.~E.~Close,
Phys. Lett. B {\bf 43}, 422 (1973).

\bibitem{FJ75}
G.~R.~Farrar and D.~R.~Jackson,
Phys. Rev. Lett. {\bf 35}, 1416 (1975).

\bibitem{MT96}
W.~Melnitchouk and A.~W.~Thomas,
Phys. Lett. B {\bf 377}, 11 (1996).

\bibitem{HR10}
R.~J.~Holt and C.~D.~Roberts,
Rev. Mod. Phys. {\bf 82}, 2991 (2010).

\bibitem{Kuh00}
S.~Kuhlmann {\it et al.},
Phys. Lett. B {\bf 476}, 291 (2000).

\bibitem{Bra12} 
L.~T.~Brady, A.~Accardi, W.~Melnitchouk and J.~F.~Owens,
JHEP {\bf 1206}, 019 (2012).

\bibitem{CJ10}
A.~Accardi, M.~E.~Christy, C.~E.~Keppel, P.~Monaghan, W.~Melnitchouk,
J.~G.~Morfin and J.~F.~Owens,
Phys. Rev. D {\bf 81}, 034016 (2010).

\bibitem{ACHL09} 
J.~Arrington, F.~Coester, R.~J.~Holt and T.~-S.~H.~Lee,
J. Phys. G {\bf 36}, 025005 (2009).

\bibitem{ARM12}
J.~Arrington, J.~G.~Rubin and W.~Melnitchouk,
Phys. Rev. Lett. {\bf 108}, 252001 (2012).

\bibitem{CJ11}
A.~Accardi, W.~Melnitchouk, J.~F.~Owens, M.~E.~Christy, C.~E.~Keppel,
L.~Zhu and J.~G.~Morfin,
Phys. Rev. D {\bf 84}, 014008 (2011).

\bibitem{CJweb}
The CTEQ-Jefferson Lab (CJ) collaboration website,
\url{http://www.jlab.org/cj}.

\bibitem{BCDMS}
A.~C.~Benvenuti {\it et al.},
Phys. Lett. B {\bf 223}, 485 (1989);
{\it ibid.} B {\bf 236}, 592 (1989).
           
\bibitem{NMCp}
M.~Arneodo {\it et al.},
Nucl. Phys. B {\bf 483}, 3 (1997).

\bibitem{NMCdop}
M.~Arneodo {\it et al.},
Nucl. Phys. B {\bf 487}, 3 (1997).

\bibitem{SLAC}  
L.~W.~Whitlow {\it et al.},
Phys. Lett. B {\bf 282}, 475 (1992).

\bibitem{Malace}
S.~P.~Malace {\it et al.},
Phys. Rev. C {\bf 80}, 035207 (2009).

\bibitem{HERA1}
F.~D.~Aaron {\it et al.},
JHEP {\bf 1001}, 109 (2010).

\bibitem{E866}  
E.~A.~Hawker {\it et al.},
Phys. Rev. Lett. {\bf 80}, 3715 (1998);
%
J.~Webb,
Ph.D. Thesis, New Mexico State University (2002),
arXiv:hep-ex/0301031;
%
P.~Reimer,
private communication.

\bibitem{CDF98}
F.~Abe {\it et al.},
Phys. Rev. Lett. {\bf 81}, 5754 (1998).    

\bibitem{CDF05}
D.~Acosta {\it et al.},
Phys. Rev. D {\bf71}, 051104(R) (2005).

\bibitem{D008}   
V.~M.~Abazov {\it et al.},
Phys. Rev. D {\bf 77}, 011106(R) (2008).

\bibitem{D0_e08} 
V.~M.~Abazov {\it et al.},
Phys. Rev. Lett. {\bf 101}, 211801 (2008).

\bibitem{CDF09}
T.~Aaltonen {\it et al.},
Phys. Rev. Lett. {\bf 102}, 181801 (2009).

\bibitem{CDFZ}
T.~Aaltonen {\it et al.}, 
Phys. Lett. B {\bf 692}, 232 (2010).

\bibitem{D0Z}
V.~M.~Abazov {\it et al.},
Phys. Rev. D {\bf 76}, 012003 (2007).

\bibitem{CDFjet1}
T.~Affolder {\it et al.},
Phys. Rev. D {\bf 64}, 032001 (2001).

\bibitem{CDFjet2}
T.~Aaltonen {\it et al.},
Phys. Rev. D {\bf 78}, 052006 (2008).

\bibitem{D0jet1}
B.~Abbott {\it et al.},
Phys. Rev. Lett. {\bf 86}, 1707 (2001).

\bibitem{D0jet2}
V.~M.~Abazov {\it et al.},
Phys. Rev. Lett. {\bf 101}, 062001 (2008). 

B.~Abbott {\it et al.},
Phys. Rev. Lett. {\bf 86}, 1707 (2001).

\bibitem{D0gamjet}
V.~M.~Abazov {\it et al.},
Phys. Lett. B {\bf 666}, 435 (2008).

\bibitem{Mason07}
D.~Mason {\it et al.},
Phys. Rev. Lett. {\bf 99}, 192001 (2007).

\bibitem{ATLAS-s}
G.~Aad {\it et al.},
Phys. Rev. Lett. {\bf 109}, 012001 (2012).

\bibitem{nCTEQ09}
I.~Schienbein, J.~Y.~Yu, K.~Kovarik, C.~Keppel, J.~G.~Morfin, F.~Olness
and J.~F.~Owens,
Phys. Rev. D {\bf 80}, 094004 (2009).

\bibitem{nCTEQ11}
K.~Kovarik, I.~Schienbein, F.~I.~Olness, J.~Y.~Yu, C.~Keppel, 
J.~G.~Morfin, J.~F.~Owens and T.~Stavreva,
Phys. Rev. Lett. {\bf 106}, 122301 (2011).

\bibitem{Eskola}
K.~J.~Eskola, H.~Paukkunen and C.~A.~Salgado,
JHEP {\bf 0904}, 065 (2009),

\bibitem{DSZS}
D.~de Florian, R.~Sassot, P.~Zurita and M.~Stratmann,
Phys. Rev. D {\bf 85}, 074028 (2012).

\bibitem{Kumano}
M.~Hirai, S.~Kumano and T.~-H.~Nagai,
Phys. Rev. C {\bf 76}, 065207 (2007).

\bibitem{Acc10}
A.~Accardi, F.~Arleo, W.~K.~Brooks, D.~D'Enterria and V.~Muccifora,
Riv. Nuovo Cim. {\bf 32}, 439 (2010).

\bibitem{E605}
G.~Moreno {\it et al.},  
Phys. Rev. D {\bf 43}, 2815 (1991).

\bibitem{NNPDF10}
R.~D.~Ball, L.~Del Debbio, S.~Forte, A.~Guffanti, J.~I.~Latorre,
J.~Rojo and M.~Ubiali,
Nucl. Phys. B {\bf 838}, 136 (2010).

\bibitem{ABKM09} 
S.~Alekhin, J.~Bl\"umlein, S.~Klein and S.-O.~Moch,
Phys. Rev. D {\bf 81}, 014032 (2010).

\bibitem{ABM11}
S.~Alekhin, J.~Bl\"umlein and S.-O.~Moch,
Phys. Rev. D {\bf 86}, 054009 (2012).

\bibitem{MSTW08}
A.~D.~Martin, W.~J.~Stirling, R.~S.~Thorne and G.~Watt,
Eur. Phys. J. C {\bf 63}, 189 (2009).

\bibitem{CT10}
H.-L.~Lai, M.~Guzzi, J.~Huston, Z.~Li, P.~M.~Nadolsky, J.~Pumplin
and C.-P.~Yuan,
Phys. Rev. D {\bf 82}, 074024 (2010).

\bibitem{JR09}
P.~Jimenez-Delgado and E.~Reya,
Phys. Rev. D {\bf 80}, 114011 (2009);
{\it ibid.} D {\bf 79}, 074023 (2009).

\bibitem{HERAPDF10}
F.~D.~Aaron {\it et al.},
JHEP {\bf 1001}, 109 (2010).

\bibitem{NNPDF13}
R.~D.~Ball {\it et al.},
Nucl. Phys. B {\bf 867}, 244 (2013).

\bibitem{Sig87} 
A.~I.~Signal and A.~W.~Thomas,
Phys. Lett. B {\bf 191}, 205 (1987).

\bibitem{MM96} 
W.~Melnitchouk and M.~Malheiro,
Phys. Rev. C {\bf 55}, 431 (1997).

\bibitem{PDG}
J.~Beringer {\it et al.} [Particle Data Group Collaboration],
Phys. Rev. D {\bf 86}, 010001 (2012),
{\tt http://pdg.lbl.gov}.

\bibitem{GP76}
H.~Georgi and H.~D.~Politzer,
Phys. Rev. D {\bf 14}, 1829 (1976).

\bibitem{EFP}
R.~K.~Ellis, R.~Petronzio and G.~Parisi,
Phys. Lett. B {\bf 64}, 97 (1976).

\bibitem{Kre04}
S.~Kretzer and M.~H.~Reno,
Phys. Rev. D {\bf 69}, 034002 (2004).

\bibitem{AQ08} 
A.~Accardi and J.-W.~Qiu,
JHEP {\bf 0807}, 090 (2008).

\bibitem{AHM09} 
A.~Accardi, T.~Hobbs and W.~Melnitchouk,
JHEP {\bf 0911}, 084 (2009).

\bibitem{Sch08} 
I.~Schienbein {\it et al.},
J. Phys. G {\bf 35}, 053101 (2008).

\bibitem{Bra11} 
L.~T.~Brady, A.~Accardi, T.~J.~Hobbs and W.~Melnitchouk,
Phys. Rev. D {\bf 84}, 074008 (2011)
[Erratum-ibid.\ D {\bf 85}, 039902 (2012)].

\bibitem{TungTMC}
K.~Bitar, P.~W.~Johnson and W.~K.~Tung,
Phys. Lett. B {\bf 83}, 114 (1979);
%
P.~W.~Johnson and W.~K.~Tung,
Print-79-1018 (Illinois Tech),
{\it Contribution to Neutrino '79, Bergen, Norway} (1979).

\bibitem{Ste06} 
F.~M.~Steffens and W.~Melnitchouk,
Phys. Rev. C {\bf 73}, 055202 (2006).

\bibitem{Ste12} 
F.~M.~Steffens, M.~D.~Brown, W.~Melnitchouk and S.~Sanches,
arXiv:1210.4398 [hep-ph].

\bibitem{DGP77}
A.~De R\'ujula, H.~Georgi and H.~D.~Politzer,
Phys. Rev. D {\bf 15}, 2495 (1977).

\bibitem{DGPannals}
A.~De R\'ujula, H.~Georgi and H.~D.~Politzer,
Ann. Phys. {\bf 103}, 315 (1977).

\bibitem{Vir92}
M.~Virchaux and A.~Milsztajn,
Phys. Lett. B {\bf 274}, 221 (1992).

\bibitem{AKL03}
S.~I.~Alekhin, S.~A.~Kulagin and S.~Liuti,
Phys. Rev. D {\bf 69}, 114009 (2004).

\bibitem{BB08}
J.~Bl\"umlein and H. B\"ottcher,
Phys. Lett. B {\bf 662}, 336 (2008).

\bibitem{Blu12}
J.~Bl\"umlein,
arXiv:1208.6087 [hep-ph].

\bibitem{MSToff}
W.~Melnitchouk, A.~W.~Schreiber and A.~W.~Thomas,
Phys. Rev. D {\bf 49}, 1183 (1994).

\bibitem{KPW94}
S.~A.~Kulagin, G.~Piller and W.~Weise,
Phys. Rev. C {\bf 50}, 1154 (1994).

\bibitem{KP06}
S.~A.~Kulagin and R.~Petti,
Nucl. Phys. A {\bf 765}, 126 (2006).

\bibitem{KMK09}
Y.~Kahn, W.~Melnitchouk and S.~A.~Kulagin,
Phys. Rev. C {\bf 79}, 035205 (2009).

\bibitem{AV18}
R.~B.~Wiringa, V.~G.~J.~Stoks and R.~Schiavilla,
Phys. Rev. C {\bf 51}, 38 (1995).

\bibitem{CDBonn}
R.~Machleidt,
Phys. Rev. C {\bf 63}, 024001 (2001).

\bibitem{WJC}
F.~Gross and A.~Stadler,
Phys. Rev. C {\bf 78}, 014005 (2008);
{\it ibid.} C {\bf 82}, 034004 (2010).

\bibitem{GL92} 
F.~Gross and S.~Liuti,
Phys. Rev. C {\bf 45}, 1374 (1992).

\bibitem{MSTplb}
W.~Melnitchouk, A.~W.~Schreiber and A.~W.~Thomas,
Phys. Lett. B {\bf 335}, 11 (1994).

\bibitem{MSS97}
W.~Melnitchouk, M.~Sargsian and M.~I.~Strikman,
Z. Phys. A {\bf 359}, 99 (1997).

\bibitem{Marquardt}
D.~W.~Marquardt,
J. Soc. Ind. Appl. Math. {\bf 11}, 431 (1963).

\bibitem{Bevington}
P.~R.~Bevington,
``Data Reduction and Error Analysis for the Physical Sciences''
(McGraw-Hill, 1969).

\bibitem{Cteq6}
J.~Pumplin {\it et al.},
JHEP {\bf 0207}, 012 (2003).
 
\bibitem{MRST02}
A.~D.~Martin, R.~G.~Roberts, W.~J.~Stirling and R.~S.~Thorne,
Eur. Phys. J. C {\bf 28}, 455 (2003).

\bibitem{CTEQweb}
The CTEQ collaboration website,
\url{http://www.cteq.org}.

\bibitem{MMSTWW}
A.~D.~Martin, A.~J.~Th.~M.~Mathijssen, W.~J.~Stirling, R.~S.~Thorne, 
B.~J.~A.~Watt and G.~Watt,
arXiv:1211.1215 [hep-ph].

\bibitem{Acc11} 
A.~Accardi,
AIP Conf. Proc. {\bf 1369}, 210 (2011).

\bibitem{Cteq61}
D.~Stump {\it et al.},
JHEP {\bf 0310}, 046 (2003).

\bibitem{LHCb}
J.~Anderson [LHCb Collaboration],
arXiv:1109.3371 [hep-ex].

\bibitem{directgamma}
D.~d'Enterria and J. Rojo, Nucl. Phys. {\bf B860}, 311 (2012), 
arXiv:1202.1762[hep-ph].

\bibitem{Lia04} 
Y.~Liang {\it et al.} [Jefferson Lab Hall C E94-110 Collaboration],
arXiv:nucl-ex/0410027.

\bibitem{Mon12} 
P.~Monaghan, A.~Accardi, M.~E.~Christy, C.~E.~Keppel, W.~Melnitchouk
and L.~Zhu,
arXiv:1209.4542 [nucl-ex].

\bibitem{AFTER@LHC}
S.~J.~Brodsky, F.~Fleuret, C.~Hadjidakis and J.~P.~Lansberg,
Phys. Rept. (2012) in press,
\url{http://dx.doi.org/10.1016/j.physrep.2012.10.001}.

\bibitem{MARATHON}
Jefferson Lab Experiment C12-10-103 [MARATHON],
G.~G.~Petratos, J.~Gomez, R.~J.~Holt and R.~D.~Ransome,
spokespersons.
 
\bibitem{BONUS12}
Jefferson Lab Experiment E12-10-102 [BONUS12],
S.~B\"ultmann, M.~E.~Christy, H.~Fenker, K.~Griffioen, C.~E.~Keppel,
S.~Kuhn and W.~Melnitchouk,
spokespersons.
 
\bibitem{SOLID}
Jefferson Lab Experiment E12-10-007 [SoLID],
P.~Souder, spokesperson.

\end{thebibliography}
\end{document}